\documentstyle[12pt] {article}
\begin {document}
\parindent=15pt
\begin{center}
\vskip 1.5 truecm
{\bf HEAVY QUARK PRODUCTION IN PARTON MODEL AND IN QCD}\\
\vspace{.5cm}
M.G.Ryskin$^1$, Yu.M.Shabelski$^2$ and A.G.Shuvaev$^1$ \\
\vspace{.5cm}
Departamento de Fisica de Particulas, Universidade de Santiago de
Compostela, \\
15706-Santiago de Compostela, Spain \\
\end{center}
\vspace{1cm}
\begin{abstract}
The cross sections of heavy quark production in $pp$ collision and for their
photo- and electroproduction are calculated in the framework of QCD. The
virtual nature of the interacting gluons as well as their transverse motion
and different polarizations are taken into account. The obtained cross
sections exhibit more rapid growth with the initial energy than the parton
model predictions and the $p_T$ distributions of produced heavy quarks
are more smooth.
\end{abstract}
\vspace{3cm}

E-mail RYSKIN@LNPI.SPB.SU  \\

E-mail SHABEL@VXDESY.DESY.DE \\

E-mail SHUVAEV@LNPI.SPB.SU \\

1 - Petersburg Nuclear Physics Institute, \\
Gatchina, St.Petersburg 188350 Russia \\

2 - Permanent address: Petersburg Nuclear Physics Institute, \\
Gatchina, St.Petersburg 188350 Russia \\

June 1995 \\

{\bf US-FT/11-95}\\

\newpage

\section{INTRODUCTION}
The investigation of the production of heavy quarks in high energy hadron
processes provides a method for studying the internal structure of hadrons.
Realistic estimates of the cross section of the heavy quark production are
necessary in order to plan experiments on existing and future accelerators.
The predicted cross section values are usually obtained in the parton model
framework and depend significantly on the quark and gluon structure functions.
These functions are practically unknown experimentally for very small values
of Bjorken variable $x$. However it is just the region that dominates in
the heavy quark production at high energies  since the main contribution to
heavy quark hadroproduction cross section comes from the $x$ values of the
order of $x\sim2m_T/\sqrt{s}\,\,$, where $m_T=\sqrt{m^2 + p^2_T}$ is
transverse mass of the produced quark.

The Gribov-Lipatov-Altarelli-Parisi (GLAP) evolution equation is used usually
to calculate the structure functions. It sums up in leading logarithm
approximation (LLA) all the QCD diagram contributions proportional
to $(\alpha_s\ln\,q^2)^n$ but it does not take into account the terms
proportional to  $(\alpha_s\ln\,1/x)^n$, so this approximation does not give
the correct asymptotic behaviour of the structure function in the small $x$
region. For the correct description of these phenomena not only the terms of
the form $(\alpha_s\ln\,q^2)^n$ have to be collected in the Feynman diagrams
but also the terms $(\alpha_s\ln\,1/x)^n$ and
$(\alpha_s\ln\,q^2\,\cdot\,\ln\,1/x)^n$.

Another problem that appears at $x\sim0$ is that of the absorption
(screening) corrections which must stop the increase of the cross section
when $x\rightarrow 0$ in accordance with the unitarity condition. It can be
interpreted as saturation of the parton density. For relatively small
virtuality $q^2\leq\,q_0^2 (x)$ the gluon structure function behaves as
$xG(x,q^2)\sim q^2R^2$, so that the cross section for the interaction of a
point-like parton with the target, $\sigma\sim(1/q^2)xG(x,q^2)\sim R^2$,
obeys the unitarity condition. The quantity $q_0^2(x)$ can be treated as a
new infrared-cutoff parameter which plays the role of a typical transverse
momentum of partons in the parton cascade of the hadron in semihard
processes. The behaviour of $q_0(x)$ was discussed in ref. [6]. It increases
with $\log(1/x)$ and at $x = 0.01-0.001$ the values of $q_0(x)$ are about
2-4 GeV.

The predictions for the cross section value of heavy quark pair production are
based usually on the parton model calculations [1, 2]. In this model all
particles involved are assumed to be on mass shell, having only longitudinal
component of the momentum and the cross section is averaged over two
transverse polarizations of the gluons. The virtualities $q^2$ of the
initial partons are taken into account only through their densities. The
latter are calculated in LLA through GLAB evolution equation. The
probabilistic picture of noninteracting partons underlies this way of
proceeding. In the region where the transverse mass $m_T$ of the produced
heavy quark $Q$ is close to $q_0(x)$ the dependence of the amplitude of the
subprocesses $gg \rightarrow \overline{Q} Q$ or $\gamma g \rightarrow
\overline{Q} Q$ on the virtualities and polarizations of the gluons should
be taken into account, i.e., the matrix elements of these subprocesses
should be calculated more accurately than is usually done in the parton
model. The matrix elements of the QCD subprocesses accounting for the
virtualities and polarizations of the gluons are very complicated. We
present here only the results for main and simplest subprocess in each case,
$gg \rightarrow \overline{Q} Q \; (\sim \alpha_{s}^{2})$ for
hadroproduction and
$\gamma g \rightarrow \overline{Q} Q \; (\sim \alpha_{s})$ for photo- and
electroproduction. The
contributions of high-order subprocesses can be essential but
our aim is to discuss the qualitative difference between our results
and the parton model predictions that can be done on the level of low-order
diagrams.

In this paper we calculate the cross section for three kinds of processes:
for the photoproduction of heavy quarks, for electroproduction (in deep
inelastic scattering) and for their production in hadron-hadron collisions.
The latter case was concidered in our previous paper \cite{3}, however the
structure function we used there was too rough numerically although it
reproduced properly all the qualitative features of the semihard
interactions. In the present paper we use two gluon structure functions. The
first one is Morfin-Tung (MT S-DIS) \cite{4} parametrization which fits
experimental data with high accuracy. The second was obtained in \cite{5}
using the same MT S-DIS boundary condition (at $Q^2 = Q^2_0$) with
accounting for the shadow corrections. We correct also a numerically small
error in the matrix element presented in \cite{3}.

\section{CROSS SECTIONS OF HEAVY FLAVOUR PRODUCTION IN QCD}

The cross section of heavy quarks photo and electroproduction is given
schematically by the graphs in Fig.1 while Fig.2 presents the sum of
diagrams for their production in $pp$ collision.
The main contribution to the cross section at small $x$ is known to come from
gluons. The lower and
upper (for $pp$ case) ladder blocks present the two-dimensional
gluon distribution $\varphi(x,q_1^2)$ and  $\varphi(x,q_2^2)$ which are
the functions of the fraction ($x$ and $y$) of the longitudinal
momentum of the initial hadron and the gluon virtuality.
Their distribution over $x$ and transverse momenta $q_T$ in hadron
is given in semihard theory \cite{6} by function $\varphi(x,q^2)$. It
differs from the usual function $G(x,q^2)$:
\begin{equation}
\label{xg}
xG(x,q^2) \,=\, \frac{1}{4\sqrt{2}\,\pi^3}
\int^{q^2}_0 \varphi (x,q^2_1)\,dq_1^2.
\end{equation}
Such definition of $\varphi(x,q^2)$ makes possible to treat
correctly the effects arising from gluons virtualities. The exact expression
for
this function can be obtained as a solution of the evolution equation which,
contrary to the parton model case, is nonlinear due to interactions between the
partons in small $x$ region.

In what follows we
use Sudakov decomposition for quarks' momenta $p_{1,2}$ through the momenta
of colliding hadrons  $p_A$ and $p_B\,\, (p^2_A = p^2_B \simeq 0)$  and
transverse ones $p_{1,2T}$:
\begin{equation}
\label{1}
p_{1,2} = x_{1,2} p_B + y_{1,2} p_A + p_{1,2T}.
\end{equation}
The differential cross sections of heavy quarks photo-, electro-\footnote{
$\sigma_{el,T,L}$ denotes the virtual photon-proton transverse and
longitudinal cross sections.} and hadroproduction have the form:\footnote{We
put the argument of $\alpha_S$ to be equal to gluon virtuality, which is very
close to the BLM scheme\cite{blm};(see also \cite{lrs}).}
$$ \frac{d\sigma_{ph}}{dy^*_1dy^*_2 d^2 p_{1T}d^2p_{2T}}\, = \,
\frac{\alpha_{em}e^2_Q}{(2\pi)^4}\frac{1}{(s)^2} \int\, d^2 q_{2T} \delta
(q_{2T} - p_{1T} - p_{2T})\, \delta(y_1 + y_2 - 1) $$ \begin{equation}
\label{sph}
\times\,\, \frac{\alpha_s (q^2_2)}{q^2_2}
\varphi (q^2_2, x)\vert M_{ph}\vert^2 \frac{1}{y_2},
\end{equation}
$$ \frac{d\sigma_{el,T,L}}{dy^*_1 dy^*_2 d^2 p_{1T}d^2
p_{2T}}\,=\,\frac{\alpha_{em}e^2_Q}{(2\pi)^4}
\frac{1}{(s)^2}\int\,d^2 q_{2T} \delta (q_{2T} - p_{1T} - p_{2T})\,
\delta(y_1 + y_2 - 1)$$
\begin{equation}
\label{sel}
\times\,\, \frac{\alpha_s (q^2_2)}{q^2_2}
\varphi (q^2_2, x)\vert M_{el,T,L}\vert^2\frac{1}{y_2},
\end{equation}
$$ \frac{d\sigma_{pp}}{dy^*_1 dy^*_2 d^2 p_{1T}d^2
p_{2T}}\,=\,\frac{1}{(2\pi)^8}
\frac{1}{(s)^2}\int\,d^2 q_{1T} d^2 q_{2T} \delta (q_{1T} +
q_{2T} - p_{1T} - p_{2T}) $$
\begin{equation}
\label{spp}
\times\,\,\frac{\alpha_s(q^2_1)}{q_1^2} \frac{\alpha_s (q^2_2)}{q^2_2}
\varphi(q^2_1,y)\varphi (q^2_2, x)\vert M_{pp}\vert^2.
\end{equation}
Here $s = 2p_A p_B\,\,$, $q_{1,2T}$ are the gluons' transverse momenta and
$y^*_{1,2}$  are the quarks' rapidities in the photon-, $\gamma^{\star}$- or
hadron-hadron c.m.s. frames,
\begin{equation}
\label{xy}
\begin{array}{crl}
x_1=\,\frac{m_{1T}}{\sqrt{s}}\, e^{-y^*_1}, &
x_2=\,\frac{m_{2T}}{\sqrt{s}}\, e^{-y^*_2},  &  x=x_1 + x_2\\
y_1=\, \frac{m_{1T}}{\sqrt{s}}\, e^{y^*_1}, &  y_2 =
\frac{m_{2T}}{\sqrt{s}}\, e^{y^*_2},  &  y=y_1 + y_2. \end{array}
\end{equation}

$\vert M_{ph}\vert^2$, $\vert M_{el,T,L}\vert^2$ and
$\vert M_{pp}\vert^2$ are the squares of the matrix elements for the cases
of photo-, electro- and hadroproduction, respectively, and indices $T$ and
$L$ denotes the transverse and longitudinal parts.

In LLA kinematic
\begin{equation}
\label{q1q2}
\begin{array}{crl}
q_1 \simeq \,yp_A + q_{1T}, & q_2 \simeq \,xp_B + q_{2T}.
\end{array}
\end{equation}
so
\begin{equation}
\label{qt}
\begin{array}{crl}
q_1^2 \simeq \,- q_{1T}^2, & q_2^2 \simeq \,- q_{2T}^2.
\end{array}
\end{equation}
(The more accurate relations are $q_1^2 =- \frac{q_{1T}^2}{1-y}$,
$q_2^2 =- \frac{q_{2T}^2}{1-x}$ but we are working in the kinematics
where $x,y \sim 0$).

The matrix elements $M$ are calculated in the Born order of QCD
without standart simplifications of the parton model since in the small
$x$ domain there are no grounds for neglecting the transverse momenta
of the gluons $q_{1T}$ and $q_{2T}$ in comparision with the quark mass
and the parameter $q_0(x)$. In the axial gauge $p^\mu_B A_\mu = 0$  the
gluon propagator takes the form $D_{\mu\nu} (q) = d_{\mu\nu} (q)/q^2,$
\begin{equation}
\label{prop}
d_{\mu\nu}(q)\, =\, \delta_{\mu\nu} -\, (q^\mu p^\nu_B \,+\, q^\nu p^\mu_B
)/(p_B q).
\end{equation}
For the gluons in $t-$channel the main contribution comes from the so called
'nonsense' polarization  $g^n_{\mu\nu}$, which can be picked out by decomposing
the numerator into longitudinal and transverse parts:
\begin{equation}
\label{trans}
\delta_{\mu\nu} (q)\, =\, 2(p^\mu_B p^\nu_A +\, p^\mu_A p^\nu_B )/s\, +\,
\delta^T_{\mu\nu} \approx\, 2p^\mu_B p^\nu_A /s\,\equiv\, g^n_{\mu\nu}.
\end{equation}
The other contributions are suppressed by the powers of $s$.
Since the sums of the diagrams in Fig.1 and Fig.2 are gauge invariant in
the LLA, the transversality condition for the ends of gluon line enables one
to replace  $p^\mu_A$  by  $-q^\mu_{1T}/x$  in expression for
$g^n_{\mu\nu}$.  Thus we get
\begin{equation}
\label{trans1}
d_{\mu\nu} (q)\,\, \approx\,\, -\,2\, \frac{p^\mu_B q^\nu_T}{x\,s}
\end{equation}
or
\begin{equation}
\label{trans2}
d_{\mu\nu} (q)\,\, \approx\,\, \,2\, \frac{q^\mu_T q^\nu_T}{xys}
\end{equation}
if we do such a trick for vector  $p_B$  too. Both these equations for
$d_{\mu\nu}$  can be used but for the form (\ref{trans1})
one has to modify the gluon
vertex slightly (to account for several ways of gluon emission --- see
ref.\cite{3} ):
\begin{equation}
\label{geff}
\Gamma_{eff}^{\nu} =
\frac{2}{xys}\,[(xys - q_{1T}^2)\,q_{1T}^{\nu} - q_{1T}^2
q_{2T}^{\nu} + 2x\,(q_{1T}q_{2T})\,p_B^{\nu}].
\end{equation}
As a result the colliding gluons can be treated as aligned ones and their
polarization vectors are directed along the transverse momenta. Ultimately,
the nontrivial azimuthal correlations must arise between the transverse
momenta $p_{1T}$ and $p_{2T}$ of the heavy quarks.

In the case of photoproduction the alignment of the gluon makes  the
distribution of the difference of the transverse momenta of the quarks
$p_T = (p_1 - p_2)_T /2$ to be proportional to $1 + cos^2 \theta$,
where $\theta$ is the angle between the vectors $p_T$ and
$q_T = (p_1 + p_2)_T /2$. In the case of hadron interactions the
polarizations of both gluons have to be taken into account as the
correlation is more complicated.

Although the situation considered here seems to be quite opposite to the
parton model there is a certain limit in which our formulae can be
transformed into parton model ones. Let us consider the $pp$ case and assume
now that the characteristic values of quarks' momenta $p_{1T}$  and
$p_{2T}$  are much larger than the values of gluons' momenta $q_{1T},q_{2T}$
\begin{equation}
\label{par1}
<p_{1T}> \gg <q_{1T}> \;, \;\; <p_{2T}> \gg <q_{2T}>
\end{equation}
and one can keep only lowest powers of $q_{1T}, q_{2T}$. It means that we can
put  $q_{1T} = q_{2T} = 0$ everywhere in the matrix element $M$ except the
vertices. Introducing the polar coordinates
\begin{equation}
\label{pol}
d^2q_{1T}\, =\, \frac{1}{2}\, dq^2_{1T} d\theta_1
\end{equation}
(and the same for $q_{2T}$) and performing angular integration with the help
of the formula
\begin{equation}
\label{aint}
\int^{2\pi}_0 d\theta_1 q^\mu_{1T} q^\nu_{1T} \, =\,\, \pi\, q^2_{1T}
\delta^{\mu\nu}_T
\end{equation}
we obtain
\begin{equation}
\label{part}
\int^{2\pi}_0\, d\theta_1 \frac{q^\mu_{1T}}{y}\frac{q^\nu_{1T}}{y}\,
\int^{2\pi}_0\, d\theta_2 \frac{q^\lambda_{2T}}{x}\frac{q^\sigma_{2T}}{x}
M_{\mu\nu}\overline{M}_{\lambda\sigma}\,
=\,2\pi^2 \frac{q^2_{1T}
q^2_{2T}}{(x\,y)^2}\,\vert M_{part}\vert^2.
\end{equation}
Here $M_{part}$ is just the matrix element in the parton model since the result
is the same as that calculated for the real (mass shell) gluons and averaged
over transverse polarizations. Then we obtain the cross section (\ref{spp})
in the form [1, 2]:
$$ \frac{d\,\sigma}{dy_1^* dy^*_2 d^2 p_{1T}} = $$ $$=\vert
M_{part}\vert^2 \frac{1}{(\hat{s})^2} \int \frac{\alpha_s (q^2_1)
\varphi(y,q^2_{1T})}{4\,\sqrt{2}\, \pi^3}
\frac{\alpha_s(q^2_2)\varphi (x,q^2_{2T})}{4
\,\sqrt{2}\, \pi^3}\, dq^2_{1T}\,dq^2_{2T}\,=$$
\begin{equation}
\label{pcs}
=\,\frac{\alpha_s^2 (q^2)}{(\hat{s})^2}\,\vert M_{part}\vert^2\,
xG(x,q^2_2)\,yG(y,q^2_1)
\end{equation}
where $\hat{s}=xys$  is the mass square of $\overline{Q} Q$ pair.

On the other hand, the used assumption (\ref{par1}) is not fulfilled in a
more or less realistic case. The transverse momenta of produced quarks
as well as gluon virtualities (QCD scale values) should be of the order of
heavy quark masses.

\section{RESULTS OF CALCULATIONS}

The functions  $\varphi (x,q^2_2)$ and $\varphi (y,q^2_1)$ are unknown at small
values of $q^2_2$ and $q^2_1$. Therefore we rewrite
the integrals over $q_2$ in eqs. (\ref{sph}) and (\ref{sel}) in the form
$$ \int^{\infty}_{0} \, d q^2_{2T} \delta (q_{2T} - p_{1T} - p_{2T})\,
\delta(y_1 + y_2 - 1)
\frac{\alpha_s (q^2_2)}{q^2_2}
\varphi (q^2_2, x)\vert M \vert^2 \frac{1}{y_2} = $$
\begin{equation}
\label{int}
4\sqrt{2}\,\pi^3 \delta (q_{2T} - p_{1T} - p_{2T})\, \delta(y_1 + y_2 - 1)
\alpha_s (Q^2_0) xG(x,Q^2_0)
(\frac{\vert M \vert^2}{q^2_2})_{q_2\rightarrow 0} \frac{1}{y_2}
\end{equation}
$$\,+\, \int^{\infty}_{Q^2_0} \, d q^2_{2T}
\delta (q_{2T} - p_{1T} - p_{2T})\, \delta(y_1 + y_2 - 1)
\frac{\alpha_s (q^2_2)}{q^2_2}
\varphi (q^2_2, x)\vert M \vert^2 \frac{1}{y_2}  $$
where eq. (\ref{xg}) is used \footnote{Please, don't mix the values of $Q^2_0$
here and $q^2_0(x)$ which was discussed in the Introduction.} and we use the
same procedure for the integral over $q_1$ in eq. (5).

In accordance with \cite{4,5}  the value $Q^2_0$ = 4 GeV$^2$ is taken. The
effective region of integration in the last integral of eq. (19) is
restricted by the values of the matrix element which decreases fast at
$q^2_i \gg m^2_Q$ so the effective QCD scale is of the same order as in the
parton model.

Using eqs. (\ref{sph}), (\ref{sel}), (\ref{spp}) and eq. (\ref{int}) we
perform the numerical calculation of heavy quark production. Two variants of
gluon distribution are used. The first one is MT (S-DIS) distribution
\cite{4} while the second is the result of ref. \cite{5}. The last one was
obtained using MT (S-DIS) boundary condition with inclusion of the shadow
corrections changing the gluon distribution in the small $x$ region. For
heavy quark masses we used the values $m_c = 1.5$ GeV and $m_b = 4.7$ GeV
and the scale values are presented in eq.(19).

In Fig. 3 we present the cross sections of charm and beauty production in
$pp$ collisions together with the lowest $\alpha_s$ order parton model
results with the same MT (S-DIS) gluon distribution. One can see that at
moderate energies where not very small $x$ values contribute, the difference
in our results with standard MT (S-DIS) gluon distribution and with the
distribution of ref. \cite{5} is negligible. An effect of screening
corrections which decrease the growth of cross sections can be seen only in
the case of charm production at LHC energies. Our cross sections are a
little higher than the parton model predictions, especially in the case of
beauty production and have more strong energy dependence.

Some time at not too high energies, where the essential $x_i$ are not small
enough the dashed curves (with the screening corrections) go higher than the
solid ones. The explanation is that after the screening corrections were
taken into account in the evolution equation \cite{5}, the values of
$\varphi (x, q^2)$ were slightly changed in comparison with the original
MT (S-DIS) set.

Our predictions for the transverse momentum distributions of produced charm
and beauty quarks are presented in Fig. 4. Here also we can see the screening
effect for $c$-quark production at $\sqrt{s}$ = 16 TeV. At large $p_T$ and
not very high energies when comparatively large $x$-values contribute there
is a small difference in cross sections obtained with gluon distributions
from \cite{4} and \cite{5}. It is the consequence of a small difference in
the values of $\varphi (x, q^2)$.

In fig. 5 the results for beauty production in $\overline{p}p$ collisions are
compared with the parton model predictions and with the existing experimental
data at $\sqrt{s}$ = 1800 GeV taken from \cite{7}. Two parton model curves
are evaluated using MRSA \cite{8} set of parton structure functions. They are
shown as dash-dotted lines. The parameters $m_b$ = 4.5 GeV, $\mu = \mu_0/2$
($\mu_0 = \sqrt{m_b^2+p_T^2}$), $\Lambda_5$ = 300 MeV and $m_b$ = 5 GeV,
$\mu = 2\mu _0$, $\Lambda_5$ = 151 MeV were used for the upper and lower
dash-dotted curves, respectively. Both our curves (solid and dashed ones)
calculated in LO are more smooth and pass between two curves of the parton
model where NLO corrections was accounted for. Our curves are in good
agreement with D0 data and with CDF data starting from $p_T >$ 10 GeV/c.
In all cases our approach results at high energies in slightly more broader
$p_T$-distributions in comparison with the parton model predictions using the
same quark mass and scale values.

Our predictions for rapidity ($y$) distributions of charm and beauty quarks
produced in $\overline{p}p$ collisions at $\sqrt{s}$ = 1800 GeV are
presented in Fig. 6. Small screening effects can be seen at large values
of $\vert y \vert$.

The total cross sections of charm and beauty quark photoproduction are
presented in Fig. 7 together with the parton model predictions in LO.
Here the situation differs slightly from the case of $\overline{p}p$
collisions, our cross sections of charm production at low energies are
smaller than the parton model predictions. However the energy dependence of
our cross sections is more strong. The screening corrections are
negligibly small until the maximal energies of HERA.

Our predictions for $p_T$ and rapidity dependences of charm and beauty quark
photoproduction at $\sqrt{s}$ = 200 GeV are shown in Figs.8 and 9.

We calculated also the transversal and longitudinal cross sections of charm
and beauty quark electroproduction in kinematical region of HERA. They are
presented in terms of the functions $F_2(x,Q^2)$ in Fig. 10. The screening
effects here are very small too. The ratios $\sigma_{L}/\sigma_{T}$,
presented in Table 1 increase with $Q^2$ (as for small $Q^2 \rightarrow 0$
$\sigma_{L}$ should tend to zero) but never exceed the value 0.4.

To study the role of small $q^2_2$ {gluon virtuality} we have changed the
value of $Q^2_0$ in eq. (19) from 4 GeV$^2$ to 1 GeV$^2$. In this case (when
the parton model formulae were used only for $q^2_2 <$ 1 GeV$^2$, while the
true $q_{2T}$ dependence of the matrix element $\vert M \vert$ was taken
into account beginning from $q^2_2 \geq$ 1 GeV$^2$) the DIS cross section
(i.e. $F_2$ values) for charm electroproduction were increased by
$\approx 5-7$\% for $W$ = 200 GeV and $Q^2$ = 4 - 64 GeV$^2$. Thus the
difference between our calculations and the parton model results increases.

Let us note that even in the region of $Q^2$ = 16 - 64 GeV$^2$ that is much
larger than the charm quark mass square $m_Q^2$ = 2.25 GeV$^2$ we observe in
Fig. 10a strong scaling violation for the function $F_2$. It reflect the
fact that the essential transverse momenta of the produced quarks $p_{Ti}$
are smaller than $\sqrt{Q^2}$ and  up to $Q^2 \approx$ 64 GeV$^2$ they stay
comparable with $m_Q$ value. Thus the scaling behaviour of $F_2$ will be
restored only at very large values of $Q^2$.

\section{CONCLUSION}

One can see that the accounting for the virtual nature of the interacting
gluons as well as their transverse motion and different polarizations result
in some qualitative differences with parton model predictions. However the
numerical values of these differences at the considered energies are possibly
smaller than the uncertainties of the predictions. Some screening effects
are expected to be essential only in the case of charm production at LHC
energies.

Our calculations of heavy quarks production based on the lowest order
matrix elements for the elementary subprocesses since our goal was to
incorporate the effects of gluon virtualities and polarizations which are
important at small $x$. The results obtained show this effects to be
actually existing. The high order corrections can be accounted for by using
the effective value of $K$-factor taken from the parton model calculations.
As it is known the main ($\sim O(\pi\alpha_S)$) part of the $K$-factor comes
from the Sudakov double log formfactors, which are the same in our scheme as
well as in the parton model.
We can expect therefore our results are to be rescaled by approximately the
same factor and the qualitative picture will be true also in this case.

This work is supported by Russian Fund of Fundamental Research
(95-2-03145).

\newpage
\begin{center}
{\bf Table 1}
\end{center}
The values of $\sigma_{L}/\sigma_{T}$ ratios for the case of charm and
beauty electroproduction at $W$ = 200 GeV.

\vskip 0.9 truecm
\begin{center}
\begin{tabular}{|c||c||c|c|} \hline

$Q\overline{Q}$ & $Q^2$ = 4 GeV$^2$ & $Q^2$ = 16 GeV$^2$
& $Q^2$ = 64 GeV$^2$  \\ \hline \hline
$c\overline{c}$   & 0.09  & 0.21  & 0.31  \\ \hline
$b\overline{b}$   & 0.012 & 0.044 & 0.13  \\  \hline

\end{tabular}
\end{center}

\newpage

\begin{center}
{\bf Figure captions}\\
\end{center}

Fig. 1. Low order QCD diagrams for heavy quark photo- and electroproduction
via photon-gluon fusion.

Fig. 2. Low order QCD diagrams for heavy quark production in $pp$ collisions
via gluon-gluon fusion.

Fig. 3. Our predictions for heavy quark production cross sections in $pp$
($\overline{p}p$) collisions together with LO order results of parton model.

Fig. 4. The calculated $p_T$ distributions of charm (a) and beauty (b) quarks
produced in high energy $pp$ ($\overline{p}p$) collisions.

Fig. 5. Our calculations of beauty production cross sections in $pp$
($\overline{p}p$) collisions at $\sqrt{s}$ = 1800 GeV (solid and dashed
curves) together with NLO order results of parton model (dash-dotted curves)
and experimental data of CDF (a) and D0 (b) Collaborations.

Fig. 6. Our predictions for rapidity distributions of heavy quark produced
in $pp$ ($\overline{p}p$) collisions at $\sqrt{s}$ = 1800 GeV.

Fig. 7. Our predictions for heavy quark photoproduction cross sections
together with LO order results of parton model.

Fig. 8. The calculated $p_T$ distributions of charm and beauty quarks
produced in $\gamma p$ interactions at $\sqrt{s}$ = 200 GeV.

Fig. 9. Our results for rapidity distributions of heavy quark produced
in $\gamma p$ interactions at $\sqrt{s}$ = 200 GeV.

Fig. 10. Our predictions for functions $F_2(x,Q^2)$ in the cases of charm
(a) and beauty (b) electroproduction in HERA kinematical region.

\end{document}